\documentclass[letterpaper, 10 pt, conference]{ieeeconf}  

\IEEEoverridecommandlockouts                              

\usepackage{graphicx}          
\usepackage{amsmath}
\usepackage{amssymb}
\usepackage{amsfonts}
\usepackage{mathrsfs}
\usepackage{comment}

\usepackage{algorithm}
\usepackage{algpseudocode}
\usepackage{xcolor}
\usepackage{tcolorbox}  
\usepackage{svg}
\usepackage{xpatch}
\usepackage{tikz}
\usepackage{comment}
\usepackage{svg}
\usepackage{parskip}
\usepackage{booktabs}

\usepackage{makecell}

\newcommand{\MK}[1]{{\color{black} #1}}

\title{\LARGE \bf
Robust Space-Filling Input Design via Stochastic Optimization*
}

\author{Máté Kiss, Roland Tóth, Maarten Schoukens
\thanks{*Funded by the European Union (ERC, COMPLETE, 101075836). Views and opinions expressed are however those of the author(s) only and do not necessarily reflect those of the European Union or the European Research Council Executive Agency. Neither the European Union nor the granting authority can be held responsible for them.}
\thanks{Máté Kiss, Maarten Schoukens and Roland Tóth are with the Control Systems Group, 5612 Eindhoven University of Technology, Eindhoven, the Netherlands. Roland Tóth is also with the Systems and Control Laboratory, Institute for Computer Science and Control, 1111 Budapest, Hungary
        (e-mail: m.kiss@tue.nl; m.schoukens@tue.nl;  r.toth@tue.nl)}%
}

\begin{document}

\maketitle
\thispagestyle{empty}
\pagestyle{empty}

\begin{abstract}
The space-filling input design approach generates a so-called space-filling dataset in the feature space of the system model. The design method is applicable on a broad class of model structures with wide selection of signals and also incorporates information measures through optimality criteria into the signal design. However, during the signal design, knowledge of a hypothesized model is required. The designed signal can perform far from the optimal if the true system is significantly different from the hypothesized system model. This paper proposes a robust space-filling input design algorithm that can generate a space-filling dataset for an entire class of models. The proposed algorithm takes the expectation of an optimality criterion over the population of the model class, and a stochastic approximation technique is employed to optimize this robust criteria. The efficiency of the proposed algorithm is demonstrated in a simulation example.

\end{abstract}

\section{INTRODUCTION}
Experiment design aims to obtain the most meaningful information from the experiment by optimally adjusting its conditions. For the identification of (nonlinear) dynamical systems from measured input-output data, the experiment design task translates into choosing the optimal input signal that returns the most informative dataset for the considered system. While data quality plays an important role for \emph{linear time-invariant} (LTI) system identification, it becomes even more important for nonlinear system identification. An LTI model may be seen as a hyperplane in the feature space, but a nonlinear model is characterized by a manifold \cite{Schoukens19}, thus being more difficult to extrapolate. Hence, a nonlinear model is much more sensitive to modeling errors and assumptions on the model structure.

Most of the optimal input design strategies are developed for LTI systems and aiming to minimize the variance of the identified model parameters using the assumption of an unbiased estimator \cite{Bombois2021, annergren2017application}. Adopting the same methodology, approaches for nonlinear systems exist only for simple cases, such as the Hammerstein and Wiener classes \cite{Colin20,Cock_Phd}, due to the simultaneous dependence on time- and frequency-domain characteristics \cite{Cock_D_optimal}. {However, as discussed previously, the leading challenge in black-box nonlinear system identification is not to have a small variance on the parameter estimates, but rather to ensure that the model is of high quality over the considered range of operation because the dominant source of errors is that the system is rarely part of the model class \cite{Schoukens19}.

In the recent years, so-called space-filling input design techniques have emerged for nonlinear systems. They create a dataset that sufficiently covers the feature space of the model \cite{herkersdorf2025online,kiss2026least,vater2024differentiable}, therefore ensuring that the identified model using this dataset behaves well over the full region of interest \cite{kiss2026least}. To determine the space-filling input, the true system has to be known \cite{hjalmarsson2005experiment,Schoukens:SysId_Freq}. Since this is often not the case, the usual practice is to design the space-filling input that is optimal for a hypothesized system model of the true system, called the nominal model.
If the system happens to differ too much from the nominal model, then the experiment performs far from optimal \cite{silvey1980optimal,box1978statistics}. Robust experiment design aims to tackle this issue by assuming uncertainty of the model parameters.
Several approaches have been proposed both in the statistical and in the control engineering literature to overcome this difficulty.

In a \textbf{sequential design}, estimation of model parameters alternates with the experiment design. Each estimation step updates the knowledge on the system parameters, and this knowledge can be used to improve the quality of the next experiment. Despite its convenience, this approach is often infeasible due to the large run-time cost of the experiment and the necessity of a correctly specified model \cite{dror2008sequential}.
A \textbf{Bayesian design} specifies a prior distribution for the model parameters and uses Bayesian inference to obtain a posterior distribution, from which an objective function can be constructed and used for experiment design. The robustness of the process hinges on the assumed prior and the likelihood of the parameters \cite{foster2021deep}.
The \textbf{min-max design} approach assumes that the parameters belong to an apriori known set and optimizes the worst possible performance of the experiment over that set. The approach is difficult to extend towards nonlinear systems because of the application specific tailoring \cite{rojas2007robust}. 
Finally, a \textbf{stochastic design} assumes that a prior probability density function of the parameters is known, and the criterion to be optimized is the expectation of some classical (non)robust optimality criterion over all possible values of the parameters \cite{huan2014gradient}. This procedure has not been studied in the input design literature and has been illustrated in a limited case study for experiment design.

Building on the concept of stochastic optimization introduced in \cite{huan2014gradient} and assuming that model parameters belong to a set as in \cite{rojas2007robust}, we propose a robust space-filling input design approach that ensures space-fillingess over the models defined by a model class. The approach is not limited to LTI systems and does not require application-specific tailoring. Furthermore, it allows for a flexible representation of the models in the class, because the model class is represented as a continuous distribution.

The rest of the paper structured as follows: Section~2 formulates the proposed robust space-filling approach. Afterward in Section~3, performance of the proposed robust design approach is demonstrated against a nominal design. Finally, Section~4 describes the conclusions.

\section{SPACE-FILLING INPUT DESIGN PROBLEM}
\subsection{System and Signal Class}
Consider a controllable, deterministic system model that characterizes our prior knowledge of the underlying true system. Its argument space, referred to as feature space, is represented by $z$. The system model approximates the process output $y$ for any given feature. The following discrete-time state-space representation is adopted as system model:
\vspace{-4mm}
\begin{subequations}\label{eq:systemModel}
\begin{align}
    x_{k+1} &= f(x_k, u_k), \label{eq:dataGeneratingSystem_a} \\
    y_k &= g(x_k, u_k), \label{eq:dataGeneratingSystem_b}
\end{align}
\end{subequations}
where $u_k\in\mathcal{U}\subseteq\mathbb{R}^{n_\mathrm{u}}$, $x_k\in\mathcal{X}\subseteq\mathbb{R}^{n_\mathrm{x}}$, and $y_k\in\mathcal{Y}\subseteq\mathbb{R}^{n_\mathrm{y}}$ are the input, state and output signals of the system at time instant $k \in \mathbb{Z}^+$. The feature space is defined as $z_k\!=\!\{u_k,x_k\}\!\in\!\mathcal{Z}\!\subseteq\!\mathbb{R}^{n_z}$, where $n_z\!=\!n_u\!+\!n_x$ and $\mathcal{Z}:\mathcal{U}\times\mathcal{X}$ denotes the joint input-state space. Consequently, the dataset is defined as $D_N\!\in\!\{ \mathcal{Z}\!\times\!\mathcal{X} \}^N$, where $\mathcal{X}$ is the predicted state.
The system model $f(\cdot):\mathcal{Z}\to\mathcal{X}$ and $g(\cdot):\mathcal{Z}\to\mathcal{Y}$ are smooth functions such that $f,g\in\mathcal{C}^1$ are at least once continuously differentiable w.r.t. any measures defined in their feature space $z\!\in\!\mathcal{Z}$ and bounded for all $k\!\in\!\mathbb{Z}^+$.

Space-filling input design optimizes a parameter vector $\theta\!\in\!{\Theta}$ w.r.t. an objective function that corresponds to an input sequence $u(\theta)$ of length $N\!\in\!\mathbb{Z}^{+}$ with a desired space-filling density. Let the parametrized input signal be given by:
\begin{equation}
   u(\theta): \Theta \to \mathbb{R}^{N \times n_u} ,
   \label{eq:inputSignalDefinition}
\end{equation}
and fulfill the requirement of $u(\theta)\!\in\!\mathcal{C}^1$ w.r.t. the parameter $\theta\!\in\!{\Theta}$ and it is bounded for all $k\!\in\!\mathbb{Z}^+$ discrete time steps and compact parameter set ${\Theta}\!\subseteq\!\mathbb{R}^{n_\mathrm{\theta}}$.
In practice, this means that (\ref{eq:inputSignalDefinition}) can represent a wide class of parametric input signals with a broad selection of signal parametrization. Depending on the choice of parametrization different signal constraint can be imposed as well (e.g., multisine signal parametrized by its amplitudes).
\subsection{Definition of Space-Fillingness}
\begin{figure}[t]
    \centering
    \includegraphics[width=\columnwidth]{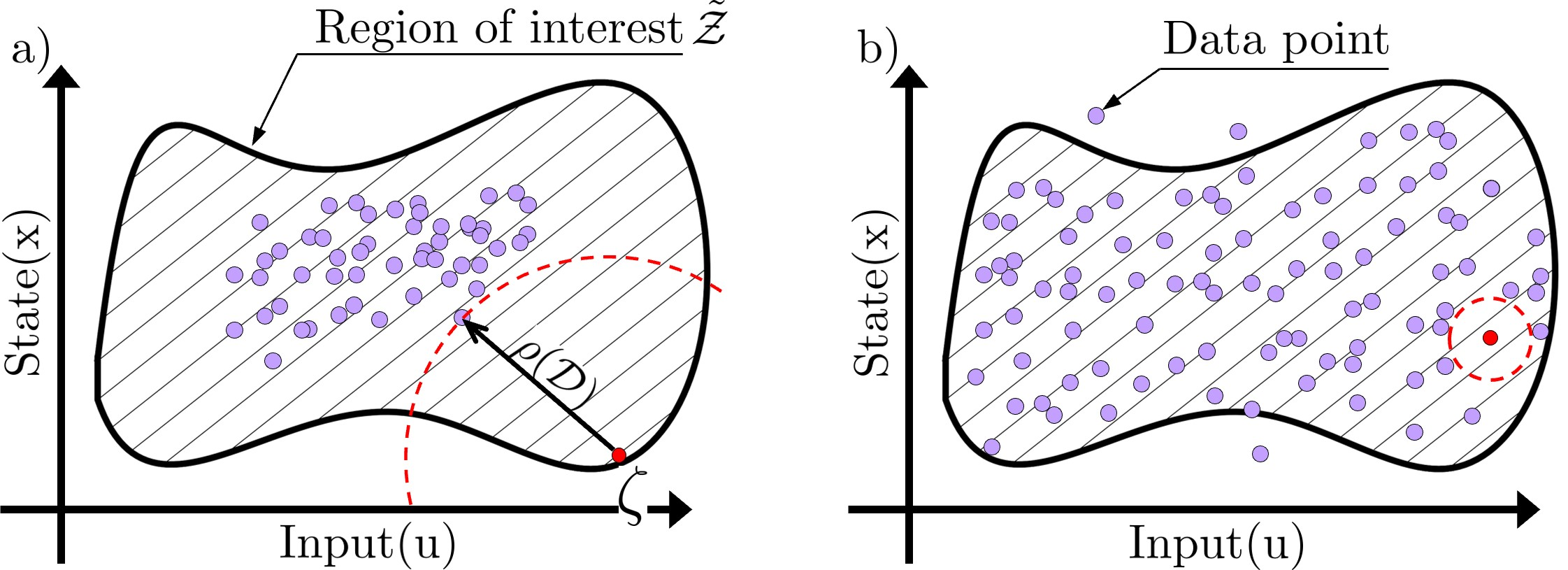}
    \vspace{-5mm}
    \caption{Space-fillingness in the feature space $\mathcal{Z}$: the striped domain denotes $\tilde{\mathcal{Z}}$ (region of interest), purple dots $D_N$ (dataset), and the red circle the largest sphere between the data points with center $\zeta$.}
    \label{fig:SpaceFillingness}
    \vspace{-6mm}
\end{figure}
Upon exciting system \eqref{eq:systemModel} with an input signal $\{u_k\}_{k=1}^{N}$, the generated dataset $D_N$ is scattered inside the region of interest $\tilde{\mathcal{Z}}$, where the region of interest is a compact subset of the joint input-state space $\tilde{\mathcal{Z}}\!\subseteq\!\mathcal{Z}$ (Fig.\ref{fig:SpaceFillingness}(a)). Then, a space-filling distance metric is defined as
\begin{equation}
\begin{gathered}\label{eq:CR}
    \rho(D_N) = \max_{\zeta \in \tilde{\mathcal{Z}}}
    \min_{\tilde{\zeta} \in D_N}
    d(\zeta, \tilde{\zeta}) ,
\end{gathered}
\end{equation}
where the radius of the largest sphere is $\min_{\tilde{\zeta} \in D_N} d(\zeta, \tilde{\zeta})$ in terms of a given distance measure $d(\cdot,\cdot):\mathcal{Z}\times\mathcal{Z}\to\mathbb{R}_0^{+}$ (e.g. Euclidean distance) between the data points without containing any other one.
The space-filling design aims to minimize $\rho(D_N)$ such that the resulting dataset is well scattered inside the region of interest, but it is not prohibited to have data point outside of it (Fig.\ref{fig:SpaceFillingness}(b)) \cite{pronzato2012design}.
\subsection{GP Regression}
GPs are universal function approximators for constructing nonparametric, probabilistic models directly from data \cite{rasmussen2003gaussian}. It allows for prior assumptions on the system to estimate an unknown nonlinear relationship $f:\mathbb{R}^{n_z}\!\to\!\mathbb{R}$ between the input $Z\!=\![z_1,\ldots,z_N]^\top\!\in\!\mathbb{R}^{N\times n_z}$ and the noise free output $X\!=\![x_1,\ldots,x_N]^\top\!\in\!\mathbb{R}^{n_x}$, generated by $x_i\!=\!f(z_{i})$. Consider the dataset $D_N\!=\!\{Z,X\}$  obtained from \eqref{eq:dataGeneratingSystem_a}. The core idea of GP-based estimation of $f$ is to consider that candidate estimates $\hat{f}$ belong to a GP, seen as a prior distribution. Then a predictive GP distribution of $\hat{f}$ is computed from the prior and the dataset $D_N$, that gives an estimate of $f$ in terms of its mean and describes the uncertainty of by its variance.

A GP $\mathcal{GP}:\mathbb{R}^n\!\to\!\mathbb{R}$ assigns to every point $z$ a random variable $\mathcal{GP}(z)$ such that for any finite set $z_1,\ldots,z_N$, the joint probability distribution of $\mathcal{GP}(z_1),\ldots,\mathcal{GP}(z_N)$ is Gaussian. Due to this property, $\hat{f}\!\sim\!\mathcal{GP}(m,\kappa)$ is fully determined by its mean $m(z)$ and its covariance function $\kappa(z,z')$, then the joint Gaussian probability is $\mathcal{N}(M_z,K_{zz})$ with $M_z\!=\![m(z_1),\ldots,m(z_N)]^\top$ and $[K_{zz}]_{i,j}\!=\!\kappa(z_i,z_j), \ i,j\!\in\!\{1,\ldots,N\}$.\\
We assume that the prior mean is zero, thus $m(z_i)\!=\!0$ and the covariance of the distribution can be well described by a squared exponential kernel:
\begin{equation}\label{eq:kernel}
    \kappa\left({  {z}},  {  {z}}'\right)\!=\!\sigma_{\mathrm{f}}^{2} \exp \left(-\frac{1}{2}( {  {z}}\!-\!  {  {z}}')^{\top} {\Lambda}^{-1}( {  {z}}\!-\! {  {z}}')\right)\!, 
\end{equation}
where $\sigma_f^2 \in \mathbb{R}$ is a scaling factor and $\Lambda = \mathrm{diag}(d_1^2,...,d_{n_\mathrm{x}}^2)$ determines the smoothness of the candidate function. Based on $D_N$ and the prior $\hat{f}\!\sim\!\mathcal{GP}(m,\kappa)$, the predictive distribution for $\hat{f}(z_\ast)$ at a test point $z_\ast$ is the posterior $\mathcal{P}(\hat{f}(z_\ast)\!\mid\!D_N,z_\ast)\!=\!\mathcal{N}(\mu(z_\ast),c(z_\ast))$ characterized by
\begin{subequations}
    \begin{align}
        \label{eq:postMean} \mu(z_\ast)\!&=\!K_Z^\top(z_\ast) K_{Z,Z}^{-1} X,\\
        \label{eq:postVar} c(z_\ast)\!&=\!\kappa(z_\ast, z_\ast)
        \!-\! K_Z^\top(z_\ast) K_{Z,Z}^{-1} K_Z(z_\ast),
    \end{align}
\end{subequations}
where $[K_Z(z_\ast)]_i\!=\!\kappa(z_i,z_\ast)$ and $K_{Z,Z}\in\mathbb{R}^{N\times N}$ with entries $[K_{Z,Z}]_{i,j} = \kappa(z_i,z_j)$ is the Gram matrix. The posterior variance $c(z_\ast)$ measures the uncertainty of the predicted model, hence a cost function is constructed based on its characteristics in the next section. Tuning of the kernel hyperparameters (i.e., $\sigma_f^2,\Lambda$) can be achieved with a wide range of methods listed in \cite{rasmussen2003gaussian}.

\subsection{V-Optimal Space-Filling Cost Function}\label{sec:V-opt cost function}
In line with the work of \cite{kiss2026least}, a latent \emph{hypothetical} model $\hat{x}_k\!=\!\hat{f}(z_k)$ can be proposed for the system \eqref{eq:systemModel}. Then, the information content of the experiment is measured in terms of the uncertainty of this \emph{hypothetical} model that would result as the posterior GP estimate based on the given data.
Upon applying the input sequence $\{u_k\}_{k=1}^N$ on the assumed model of the system \eqref{eq:systemModel}, the dataset $D_N$ can be obtained. From this dataset $D_N$, the matrix of features $Z\!\in\!\mathbb{R}^{N\times n_z}$ is constructed, where the columns $n_z$ span the space in that we would like to be space-filling. Given the features, the posterior behavior is described by $\hat{f}\!\sim\!\mathcal{GP}\left(\mu, c\right)$. The region of interest $\tilde{\mathcal{Z}}$ is discretized by $M$ number of anchor points. Thus, the anchor set is defined by $D_I = \{ \tilde{z}_i \}_{i=1}^M$ such that all elements in the set are distinct.

A space-filling promoting cost function is achieved by using a kernel \eqref{eq:kernel} that is monotonically decreasing w.r.t. $||z-z'||$ where $z$ is considered as a data point and $z'$ as an anchor point \cite{kiss2026least}. Hence, the distance between $z$ and $\tilde{z}$ decays as the distance metric $||z-\tilde{z}||$ increases. Consequently, the posterior variance of the \emph{hypothetical} model $\hat{f}$ is defined as
\begin{align}\label{eq:posteriorVariance}
   c(\tilde{z}) &= \kappa(\tilde{z}, \tilde{z})\!-\!K_Z^\top(\tilde{z}) K_{Z,Z}^{-1} K_Z(\tilde{z}) .
\end{align}
Choosing the information metric of the experiment design to be the average posterior variance  of the model prediction $\hat{f}$ over the design domain $\tilde{Z}$, translates to the so called V-optimality criterion \cite{atkinson1992}. It is expressed as the following scalar valued space-filling cost function, namely the average posterior variance evaluated at the anchor points:
\begin{equation}\label{eq:costFunction}
    \mathcal{V}(\theta):=\frac{1}{M} \sum_{i=1}^{M} c(\tilde{z}_i).
\end{equation}
Now, the space-filling input design problem can be expressed as an optimization problem to find a parameter vector $\theta$ that yields an input sequence $u_k(\theta)$ capable of generating a space-filling dataset $D_N(\theta)$ in case of applying to the system \eqref{eq:systemModel} with any given initial condition $z^{(0)}$. Minimizing the resulting cost \eqref{eq:costFunction} ensures the space-filling behavior inside the region of interest $\tilde{Z}$:
\begin{subequations}
\begin{align}
\underset{\theta\in{\Theta}}{\text{min}} 
 & \quad \mathcal{V}({\theta}) \label{eq:opt1}\\
 \text{s.t.} & \quad z_0=z^{(0)},\\
 & \quad x_{j+1} \!=\! f(z_j), \quad j\!\in\!\{0,\cdots, N \}\\
 &\quad D_N\!=\!\{z_j,x_j\}_{j=1}^{N}.
\end{align}
\end{subequations}
While standard space-filling designs aim to minimize the distance metric \eqref{eq:CR}, the GP-based approach makes use of the fact that the covariance function inherently defines a distance metric based on the data distribution. In this minimax-type of space-filling design, the decision variable $\theta$ is optimized using a space-filling cost $\mathcal{V}(\theta)$, which incorporates the covariance function and, in turn, influences $D_N$ to optimize the space-filling measure $\rho(D_N)$. For the more interested readers on the GP-based space-filling input design we refer to the work of \cite{kiss2026least}.

\section{ROBUST SPACE-FILLING INPUT DESIGN APPROACH}\label{sec:robustApproach}
This section presents the robust space-filling input design method using a stochastic optimization technique. For a dynamical nonlinear problem, \cite{huan2014gradient} showed that by assuming a population of the sought model parameters with known statistics and optimizing the expected value of a criterion over this population, can lead to a robust experiment design.

Adopting a similar perspective, let $\Xi$ be a compact model space containing models $\sigma(\eta)\!\in\!\Xi$ with distribution $P(\sigma)$. Each model $\sigma_i(\eta)$ is parametrized by the model parameter vector $\eta\!\in\mathbb{R}^{n_\eta}\!$. Then, we can define a model class representing all models, each having its own feature space, in which we want to achieve space-fillingness:
\begin{equation}\label{eq:modelClass}
    \mathcal{M} = \{ \sigma \in \Xi \} .
\end{equation}
For notational convenience we drop the depending terms and use $\sigma(\eta)\!=\!\sigma$, unless stated otherwise.

We look for an experiment minimizing the expectation of a proposed space-filling cost function $\mathcal{V}(\theta,\sigma)$
\begin{equation}
    \mathcal{R}(\theta)\!=\!\mathbb{E}_{\mathcal{M}}[ \mathcal{V}(\theta,\sigma)]\!=\!\int_{\mathcal{M}} \mathcal{V}(\theta,\sigma) P(\sigma) d\sigma,
\end{equation}
where the minimization variable are the signal parameters $\theta$ and the cost function $\mathcal{R}(\theta)$ measures the average information over all possible models. Following the stochastic optimization techniques \cite{bottou98}, we consider $P(\sigma)$ as an unknown distribution and estimate $\mathcal{R}$ with its empirical measure:
\begin{equation}\label{eq:loss}
    \mathcal{R}(\theta)\!\approx\!\hat{\mathcal{R}}_{\mathfrak{B}}(\theta)\!=\!\frac{1}{L} \sum_{i=1}^{L} \mathcal{V}(\theta,\sigma_i) ,
\end{equation}
where we consider a batch of $L$ sampled models $\mathfrak{B}\!=\!\{ \sigma_i \}_{i=1}^{L}$ from the class $\mathcal{M}$. 

The cost function \eqref{eq:loss} is minimized using batch stochastic gradient descent (SGD):
\begin{equation}\label{eq:SGD}
    \theta^{\iota+1}\!=\!\theta^\iota - \gamma^\iota \nabla_\theta \hat{\mathcal{R}}_{\mathfrak{B}}(\theta^\iota) ,
\end{equation}
where $\gamma^\iota$ is the learning rate and $\iota\!\in\!\mathbb{N}_1^N$ denotes the iteration number. Each iteration consists of randomly sampling $L$ number of models from class $\mathcal{M}$ and averaging their costs at a given $\theta^\iota$. The proposed method is summarized in Algorithm~\ref{alg:RobustSGD}.
It has been showed that under mild conditions (e.g., smoothness of the system functions \eqref{eq:systemModel}) stochastic gradient descent optimization converges to the closest local minimum \cite{bottou98}.
\begin{algorithm}[t]
\caption{Mini-batch SGD for Robust Space-Filling Input Design}
\label{alg:RobustSGD}
\begin{algorithmic}[1]
\Require Initial parameter $\theta^{(0)}$, step-size sequence $\{\gamma^\iota\}$, batch size $L$, region of interest $\tilde{\mathcal{Z}}$, anchor set $D_I$, GP hyperparameters $\Lambda$, $\sigma_{\mathrm{f}}^{2}$.
\State Initialize $u_k(\theta)$ with $\theta^{(0)}$;
\State Set iteration index $\iota \gets 0$;
\Repeat
    \State Draw a batch of models $\mathfrak{B}^\iota\!=\!\{\sigma_i\}_{i=1}^{L}$ from the model class $\mathcal{M}$;
    \State Compute the loss $\hat{\mathcal{R}}_{\mathfrak{B}}^{\iota}(\theta)\!=\!\frac{1}{L} \sum_{i=1}^{L} \mathcal{V}(\theta^{\iota},\sigma_i)$ for all models in the batch $\sigma_i\!\in\!\mathfrak{B}^\iota$ with the chosen optimality criterion $J(\theta,\sigma)$;
    \State Update
        $\theta^{\iota+1}\!=\!\theta^{\iota} - \gamma^\iota \nabla_\theta \hat{\mathcal{R}}_{\mathfrak{B}}^{\iota}(\theta^{\iota})$;
    \State Set $\iota \gets \iota + 1$;
\Until{Convergence or a termination condition is satisfied}
\State \Return $\hat{\theta}\!\gets\!\theta^{\iota}$, and adopt $u_k(\hat{\theta})$ as the robust space-filling input signal.
\end{algorithmic}
\end{algorithm}

\section{SIMULATION STUDY}
First, we set up an experiment for designing a space-filling signal for a nominal model. Next a model class, that includes the nominal model as well, is defined and a robust space-filling input is designed for this class using the proposed robust approach introduced in Sec~\ref{sec:robustApproach}. Finally, the nominal and robust designs are compared to each other in terms of space-fillingness and optimization cost on 100 randomly drawn models from the class $\mathcal{M}_1$. We also investigate how the system class distribution affects the robust design. To this end, we compare 3 different model classes ($\mathcal{M}_{1,2,3}$) to each other in terms of space-fillingness and optimization cost on 100 randomly drawn models from their respective classes.
\subsection{Nominal Model, Model Class and Region of Interest}\label{sec:SimulationStudy_NomVSRob}
Consider the nominal model as a nonlinear second order mass-spring-damper system:
\begin{equation}\label{eq:nominalModel}
\sigma(\bar{\eta})\!=\!
\left\{
\begin{aligned}
\begin{bmatrix}
\dot{x}_1 \\
\dot{x}_2
\end{bmatrix}
&=
\begin{bmatrix}
x_2 \\
\frac{1}{\bar{m}} \left(u\!-\!\bar{s}x_1\!+\!\frac{\bar{s} \bar{l} x_1}{\sqrt{x_1^2+\bar{a}^2}}\!-\!\bar{b}x_2 \right)
\end{bmatrix}
\end{aligned}
\right.
\end{equation}
where $x_1$ and $x_2$ denote the position and the velocity. Notice that this example has also been discussed in \cite{kiss2026least,vater2024differentiable} for input design. We use the following shorthand notation $\bar{\sigma}\!=\!\sigma(\bar{\eta})$ for the nominal model, unless stated otherwise. The defined model depends on the nominal parameter values $\bar{\eta}\!=\![\bar{m},\bar{s},\bar{b}, \bar{l}, \bar{a}]$ . The parameters are $\bar{m}\!=\!5$~kg, $\bar{s}\!=\!800$~N/m and $\bar{b}\!=\!10$~Ns/m for the mass, stiffness, and damping coefficient, respectively. The tensionless length of the spring is $\bar{l}\!=\!0.17$~m while the maximum stretched length is $\bar{a}\!=\!0.25$~m. 

The robust experiment considers the following model class:
\begin{equation}\label{eq:robustDesignModelClass}
    \mathcal{M}\!=\! \{ \sigma_i(\eta_i, \theta) \}_{i=1}^{\infty},
\end{equation}
where every model has the same model structure as \eqref{eq:nominalModel}. The model parameters $\eta_i\!=\![m,s,b,l,a]$ are sampled from their own continuous distributions as detailed in Table~\ref{t:Distributions}. For $\mathcal{M}_1$, all parameters are sampled from uniform distributions, with $P_m\!\sim\!\mathcal{U}(0.7\bar{m},1.3\bar{m})$, $P_s\!\sim\!\mathcal{U}(0.7\bar{s},1.3\bar{s})$, $P_b\!\sim\!\mathcal{U}(0.7\bar{b},1.3\bar{b})$ and $P_l\!\sim\!\mathcal{U}(0.9\bar{l},1.1\bar{l})$, $P_a\!\sim\!\mathcal{U}(\bar{a},1.1\bar{a})$.
The model class $\mathcal{M}_2$ uses a skew symmetric $Beta(\alpha,\beta)$ distribution, while $\mathcal{M}_3$ uses a $Beta(5,5)$ distribution which is a bounded approximation of the normal distribution. To avoid unstable models, all distributions are bounded. The bounds are determined by the upper and lower values of the defined uniform distributions of $\mathcal{M}_1$.

\begin{table}[t]
\centering
\caption{Model class distributions and obtained final costs $\hat{\mathcal{R}}_{10}(\theta)$.}
\begin{tabular}{|c||c|c|c|}
\hline
\text{$\eta$} & \rule{0pt}{2.6ex} $\mathcal{M}_1$ & $\mathcal{M}_2$ & $\mathcal{M}_3$ \\ \hline\hline
m  & $\mathcal{U}$30\% & $Beta(2,5)$ & $Beta(5,5)$\\ \hline
s  & $\mathcal{U}$30\% & $Beta(2,5)$ & $Beta(5,5)$\\ \hline
b  & $\mathcal{U}$30\% & $Beta(2,5)$ & $Beta(5,5)$\\ \hline
l  & $\mathcal{U}$10\% & $Beta(2,5)$ & $Beta(5,5)$\\ \hline
a  & $\mathcal{U}$10\% & $Beta(2,5)$ & $Beta(5,5)$\\ \Xhline{1.2pt}
\rule{0pt}{2.6ex} $\hat{\mathcal{R}}_{10}(\theta)$  & $0.12$ & $0.09$ & $0.10$ \\ \hline
\end{tabular}
\label{t:Distributions}
\vspace{-6mm}
\end{table}

The region of interest is a rectangle in the 2-dimensional state space $\mathcal{\tilde{Z}} \!=\! \{ (x_1,x_2) \!\in\! \mathbb{R}^2 \mid x_1 \!\in\! [-0.1,0.1], \; x_2 \!\in\! [-0.8,0.8] \}$. It is represented by 7 equally distanced anchor points $\tilde{z}_i$ along each dimension of the space, giving $M=49$ anchor points in total (see examples in Fig.~\ref{fig:boxPlot}). \MK{The input enters \eqref{eq:nominalModel} linearly, allowing its dimension to be omitted from the region of interest.} The kernel widths are chosen to be equal with the adjacent anchor point distance in the corresponding dimension $\Lambda = \mathrm{diag}(0.03,0.26)$ and the scaling factor is $\sigma_{\mathrm{f}}^{2} = \sqrt{10}$.

\subsection{Input Signal}\label{sec:InputSignal}
Both the nominal and the robust design starts from the same multisine signal $u\!\in\!u(\Theta)$ with parameters $\theta\!=\!\mathrm{vec}(\{\mathrm{A}_j,\varphi_j\}_{j=j_\mathrm{min}}^{j_\mathrm{max}})$:
\begin{equation}\label{eq:multisine}
    u_k(\theta) \!=\! \sum_{j=j_\mathrm{min}}^{j_\mathrm{max}} \mathrm{A}_j \sin\left(2\pi j \frac{\omega_0}{\omega_s}k + \varphi_l\right).
\end{equation}
Between the range $\omega_\mathrm{{min}}$=1~Hz and $\omega_\mathrm{{max}}$=10~Hz, every $\mathrm{7}^{th}$ frequency line is excited which corresponds to a total number of excited frequencies $J_\omega\!=\!14$ with $j_\mathrm{min}\!=\!12$, $j_\mathrm{max}\!=\!103$. The phases are initialized using a uniform random distribution $[0,2 \pi[$, while $\omega_0\!=\!\omega_s/N$ corresponds to the frequency resolution with the sampling frequency $\omega_s$=100~Hz and $N$=1024 data samples per period. 
The amplitudes $\{\mathrm{A}_j\}_{j=j_\mathrm{min}}^{j_\mathrm{max}}$ are parametrized such that  their value can change per frequency line. In our simulation study, every excited frequency starts with the initial amplitude of 8~N.

\subsection{Input Design} \label{sec:InputDesign}
First using the method introduced in Sec.~\ref{sec:V-opt cost function}, the nominal space-filling signal $u(\bar{\theta})$ is designed by minimizing cost function \eqref{eq:costFunction}, where the data is generated by the nominal model $\bar{\sigma}$. Then, a robust experiment design takes place according to Sec.~\ref{sec:robustApproach}. During the robust design, we randomly draw a batch of $L=10$ models from the model class $\mathcal{M}_1$ defined in \eqref{eq:robustDesignModelClass}. To apply the SGD algorithm \eqref{eq:SGD}, the optimization cost $\hat{\mathcal{R}}_{10}(\theta)$ is computed by evaluating the V-optimal cost \eqref{eq:costFunction} for each sampled model in the batch\footnote{\label{fn:oracleKnowledge}To compute the V-optimal cost for each randomly selected models in the batch, we require access to the model states $\{x_k\}_{k=1}^N$. These are obtained by simulating the model structure represented by \eqref{eq:nominalModel} with the sampled $\eta_i$ model parameters.}. A gradient step with step size $\gamma^\iota$ is then performed, after which a new batch of models is sampled. This procedure is repeated until convergence or a termination condition is met. The optimization is accomplished with the ADAM optimizer using gradients computed through backpropagation.
\begin{figure}[t]
    \centering
    \includegraphics[width=\columnwidth]{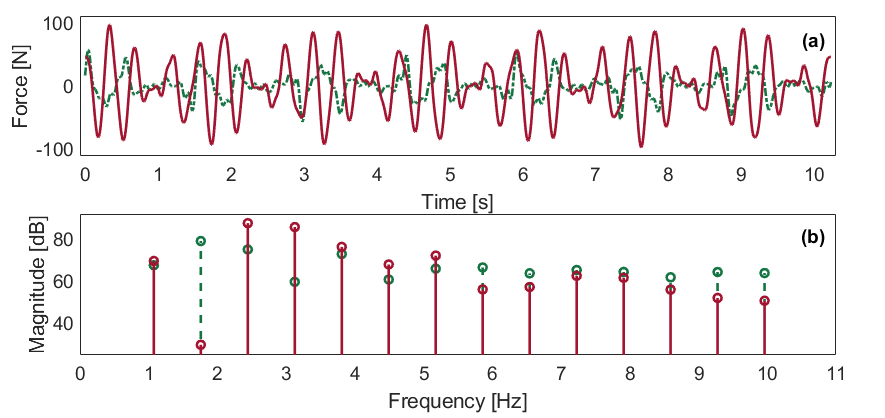}
    \vspace{-8mm}
    \caption{(a) Robust $u(\hat{\theta})$ (red) and nominal $u(\bar{\theta})$ (green) space-filling signal on the time domain, (b) excited frequencies of $u(\hat{\theta})$ (red) and $u(\bar{\theta})$ (green). Robust design is done on $\mathcal{M}_1$.}
    \label{fig:signals}
    \vspace{-3mm}
\end{figure}

\begin{figure}[t]
    \centering
    \includegraphics[width=\columnwidth]{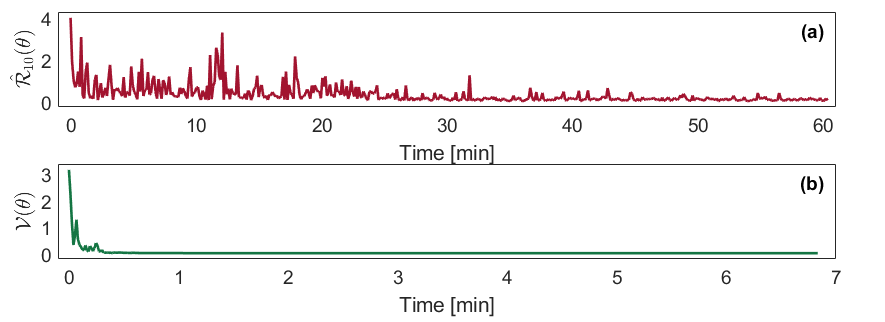}
    \vspace{-8mm}
    \caption{Evolution of (a) the robust optimization cost $\hat{\mathcal{R}}_{10}(\hat{\theta})$ and (b) the nominal design cost $V(\bar{\theta})$. Robust design is done on $\mathcal{M}_1$.}
    \label{fig:loss}
    \vspace{-6mm}
\end{figure}

Figure~\ref{fig:signals} shows the robust $u(\hat{\theta})$ and the nominal $u(\bar{\theta})$ space-filling signals on the time domain and also their excited frequencies. The nominal design converged after 6 minutes and 49 seconds with final cost value $V(\bar{\theta})\!=\!0.07$, whereas the robust design was considered to be converged after 500 iterations, which took 60 minutes and 19 seconds with final cost value $\hat{\mathcal{R}}_{10}(\hat{\theta})\!=\!0.12$ (Fig.~\ref{fig:loss}). To quantify whether the longer optimization pays off, we analyze the performance of both signals on the nominal model $\bar{\sigma}$ in the first step.
Results are summarized in the third and fourth column of Table~\ref{t:IdentificationResults}. The space-fillingness of the two signals, evaluated on the nominal model $\bar{\sigma}$, is presented in Fig.~\ref{fig:boxPlot}(c) and (d). In this particular case, the nominal signal $u(\bar{\theta})$ has a 13\% smaller covering radius and achieves 38\% lower V-optimal cost compared to the robust signal $u(\hat{\theta})$. The performance fallback of the robust design on the nominal model is expected, since the advantage of the robust design lies in its ability to account for the model variance.

Next, we draw 100 random models from the class $\mathcal{M}_1$ and excite them once with the robust $u(\hat{\theta})$ and once with the nominal signal $u(\bar{\theta})$ to compare the obtained covering radius $\mathcal{\rho}(D_N)$ and V-optimal cost $\mathcal{V}(\theta)$. The obtained statistics are written in the first and second column of Table~\ref{t:IdentificationResults}. As depicted in Fig.~\ref{fig:boxPlot}(a), the mean covering radius with the robust signal is only 25\% lower than that obtained with the nominal signal. This is due to the special system characteristics; namely, the model leaves an open gap in the middle of the region of interest. Nevertheless, this indeed indicates a better space-filling performance over the model class. The mean V-optimal cost is 4.8 times lower with the robust signal opposed to the nominal one (Fig.~\ref{fig:boxPlot}(b)).
\begin{figure}[t]
    \centering
    \includegraphics[width=\columnwidth]{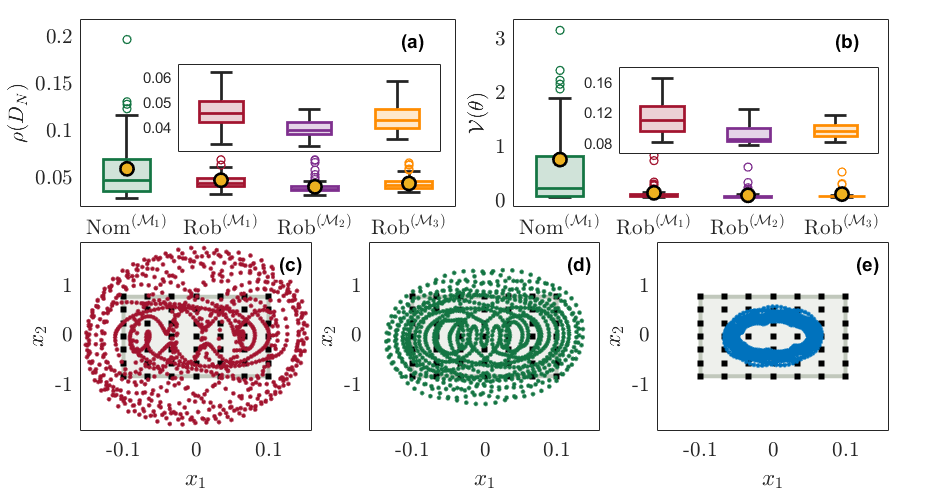}
    \vspace{-8mm}
    \caption{(a) box plots of the covering radiuses $\rho(D_N)$ over 100 randomly drawn models from different $\mathcal{M}$ with their mean value (yellow circle), (b) box plots of the V-optimality $\mathcal{V}(\theta)$ over 100 randomly drawn models from different $\mathcal{M}$ with their mean value (yellow circle), (c,d) robust $D^{(\bar{\sigma})}(\hat{\theta})$ (red) and nominal $D^{(\bar{\sigma})}(\bar{\theta})$ (green) space-filling dataset on the nominal system, (e) dataset obtained with the initial multisine signal on the nominal system. The gray area highlights the region of interest $\mathcal{\tilde{Z}}$ and the black squares mark the anchor points $D_I$.}
    \label{fig:boxPlot}
\end{figure}

\begin{table}[t]
\centering
\caption{Mean performance of the compared approaches over the model class $\mathcal{M}_1$ and on the nominal model $\bar{\sigma}$}
\begin{tabular}{|c||c|c|c|c|}
\hline
\text{Design case} & \rule{0pt}{2.6ex} $\rho^{(\mathcal{M}_1)}$ & $\mathcal{V}^{(\mathcal{M}_1)}$ & $\rho^{(\bar{\sigma})}$ & $\mathcal{V}^{(\bar{\sigma})} $ \\ \hline\hline
Robust & 0.048 & 0.163 & 0.044 & 0.115\\ \hline
Nominal  & 0.060 & 0.782 & 0.038 & 0.071 \\ \hline
\end{tabular}
\label{t:IdentificationResults}
\vspace{-6mm}
\end{table}

The observed difference between the robust and nominal designs, as reflected in the variance of the box plots, indicates that the proposed robust design approach allocates its energy more uniformly over the model class to be more robust in terms of the defined space-filling cost. It should be noted that outliers arise from models that tend to leave a large gap in the middle of the region of interest (see for example Fig.~\ref{fig:randomSamples}(b)). This behavior occurs since the considered multisine would require higher frequency content to be able to cover the central gap.

As illustrated by the three randomly drawn examples in Fig.~\ref{fig:randomSamples}, when we are close to the hypothesized optimal model $\bar{\sigma}$, the robust signal does not necessarily outperform the nominal one (Fig.~\ref{fig:randomSamples}(c)). However, as the model deviates further from $\bar{\sigma}$, the space-filling performance increases (Fig.~\ref{fig:randomSamples}(a)). \\
It has been shown in prior works that good space-filling experiments lead to better identification results \cite{kiss2026least}.
\begin{figure}[t]
    \centering
    \includegraphics[width=\columnwidth]{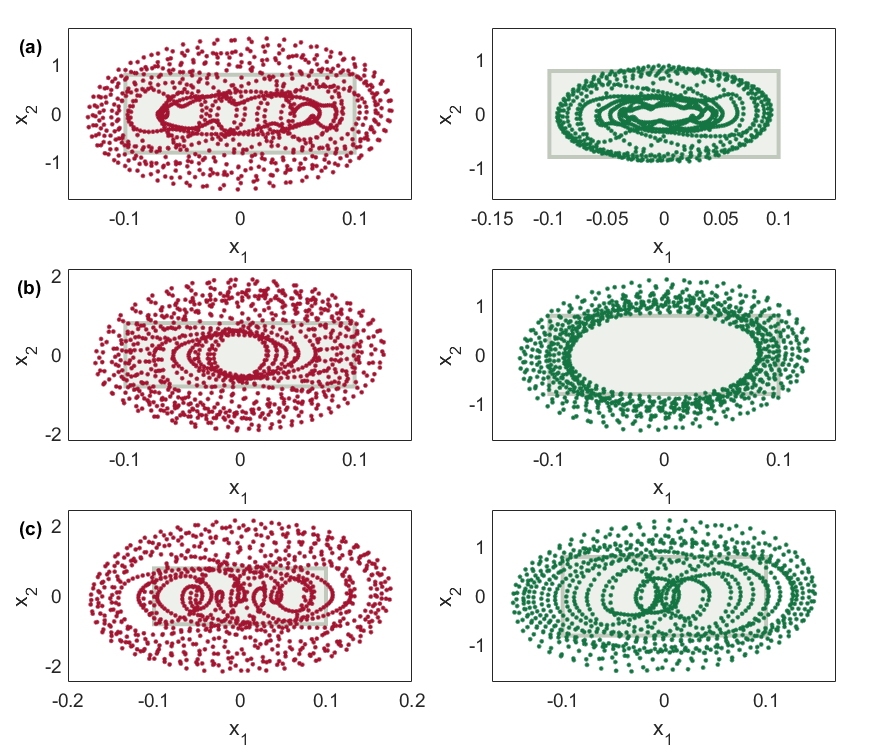}
    \vspace{-8mm}
    \caption{Obtained space-filling data sets  with the robust signal $u(\hat{\theta})$ (red) and with the nominal signal $u(\bar{\theta})$ (green) on three different randomly drawn models (a-c) from $\mathcal{M}_1$. The gray area highlights the region of interest $\mathcal{\tilde{Z}}$.}
    \label{fig:randomSamples}
    \vspace{-6mm}
\end{figure}

The effect of the model class distribution is also represented on Figure~\ref{fig:boxPlot}(a),(b). In all cases, we let the robust design algorithm (Alg.~\ref{alg:RobustSGD}) run for 500 iterations and compare the optimizations based on the observed mean V-optimal cost ($\mathcal{V}^{(\mathcal{M}_2)}\!=\!0.117$, $\mathcal{V}^{(\mathcal{M}_3)}\!=\!0.140$) and mean covering radius ($\rho^{(\mathcal{M}_2)}\!=\!0.041$, $\rho^{(\mathcal{M}_3)}\!=\!0.045$). The defined parameter distributions in $\eta$ can be seen as weights over the parameters, expressing our belief about the true value of that parameter. A uniform distribution can be seen as having no prior knowledge, resulting in a broad model class. In contrast, a skew-symmetric or normal distribution reflects some confidence about the true value of the parameter and also about our system. As a result, the algorithm samples from a more concentrated set of models, yielding to models with more similar stability characteristics. This fact facilitates less variance and a lower robust cost (Table~\ref{t:Distributions}). 

There are three tradeoffs to manage in the proposed robust space-filling input design:
(\textbf{i}) There is a tradeoff between robustness and nominal optimality. It can be mitigated by modifying the distribution of the parameters in ${\eta}$. Uniform distributions aim to provide equal performance over the full model class, while e.g., a normal distribution around a nominal parameter vector will put a stronger emphasis on a good nominal performance.
(\textbf{ii}) There is also a tradeoff between the convergence speed and optimization accuracy of the SGD. Faster decaying learning rates $\gamma$ usually yield a rapid initial decrease in the cost, but slow down convergence to the optimum. While the convergence speed is also influenced by the variance of the optimization function $\hat{\mathcal{R}}_{\mathfrak{B}}(\theta)$, which is affected by the stochastic approximation of the gradient. A larger batch size $L$ reduces the variance, at the expense of a linearly growing computational cost.
(\textbf{iii}) The final tradeoff is between the input signal parametrization and optimization accuracy. An under parametrized signal (e.g. insufficient number of excited frequencies in $\theta$) can limit both convergence speed and accuracy, as different models in the class may have substantially different signal parameter requirements.

\section{CONCLUSIONS}
In this paper, a robust space-filling input design approach has been proposed via stochastic optimization technique to create an input signal for a class of models such that it guarantees the coverage in the region of interest of each model. By optimizing the expected value of an optimality criterion over the population of the model class with the stochastic gradient approach, the generated input results in a space-filing design for all model belonging to the considered model class. As a consequence, the average coverage of the region of interest significantly increased, while the average experimental cost could be significantly reduced over the model class.




{
\parskip=0pt
\bibliographystyle{IEEEtran}
\bibliography{references}
}
\end{document}